# Funneling with the Two-Beam RFQ*

H. Zimmermann, A. Bechtold, A. Schempp, J. Thibus, IAP, Frankfurt, Germany

Institut für Angewandte Physik, Johann Wolfgang Goethe-Universität,

Robert-Mayer-Straße 2-4, D-60054 Frankfurt am Main, Germany

*Abstract*

New high current accelerator facilities like proposed for HIDIF or ESS require a beam with a high brilliance. These beams can not be produced by a single pass rf-linac. The increase in brightness in such a driver linac is done by several funneling stages at low energies, in which two identically bunched ion beams are combined into a single beam with twice the frequency current and brightness. Our Two-Beam-RFQ funneling experiment is a setup of two ion sources, a two beam RFQ, a funnel deflector and beam diagnostic equipment to demonstrate funneling of ion beams as a model for the first funneling stage of a HIIF driver. The progress of the funneling experiment and results of simulations will be presented.

## 1 INTRODUCTION

The beam currents of linacs are limited by space charge effects and the focusing and transport capability of the accelerator. Funneling is doubling the beam current by the combination of two bunched beams preaccelerated at a frequency $f_0$ with an rf-deflector to a common axis and injecting into another rf-accelerator at frequency $2*f_0$, as indicated in fig. 1.

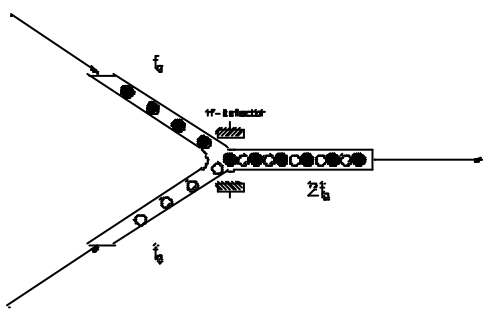

Fig 1: Principle of funneling.

By the use of the two-beam RFQ the two beams are brought very close together while they are still radially and longitudinally focused. Additional discrete elements like quadrupole-doublets and -triplets, debunchers and bending magnets, as they have been proposed in first funneling studies, are not necessary [1,2,3]. A short rf-funneling deflector will be placed around the beam crossing position behind the RFQ (fig. 2). The layout of the proposed HIDIF-injector with two-beam RFQs in front of the first and second funneling sections is shown in figure 3. The HIDIF linac starts with 16 times 3 ion sources for three different ion species to allow so-called „telescoping" at the final focus [5]. With four funneling stages the frequency has been increased from 12.5 MHz to 200 MHz accordingly [6].

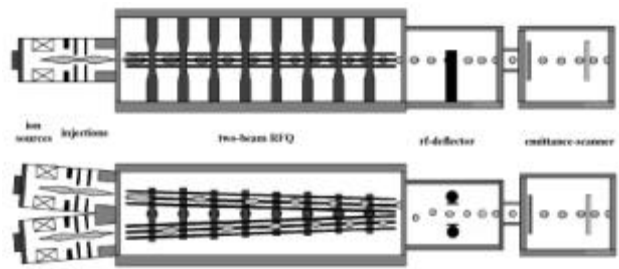

Fig. 2: Experimental set-up of the two-beam funneling experiment.

For studies of the new two-beam RFQ structure and the rf-deflector, the first two-beam funneling experiments have been done with $He^+$-ions at low energies to facilitate ion source operation and beam diagnostics.

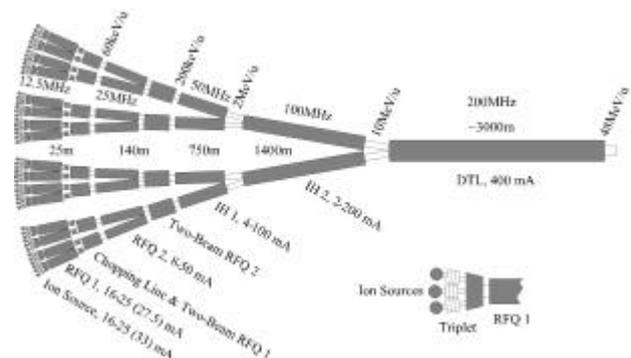

Fig. 3: Layout of the 12.5...200MHz linac system for 400 mA of $Bi^+$.

Two small multicusp ion sources and electrostatic lenses, built by LBNL (Lawrence Berkeley National Laboratory) [7,8] are used. The ion sources and injection systems are attached directly on the front of the RFQ with an angle of 76 mrad, the angle of the beam axes of the two-beam RFQ.

*Work supported by the BMBF

Figure 2 shows a scheme of the experimental set-up of the two-beam funneling experiment with a slit-grid emittance measurement device on the right side.

## 2 ION SOURCES INJECTION SYSTEMS AND TWO-BEAM RFQ

Two multicusp ion sources have to deliver two identically ion beams. This operation has been tested on an emittance measurement device. The measured emittances of both ion-sources show differences up to 30% [4].

The two-beam RFQ consists of two sets of quadrupole electrodes, where the beams are bunched and accelerated with a phase shift of 180° between each bunch, driven by one resonant structure. With the use of identical RFQ electrode designs for both beam lines, the electrodes of one beam line are installed with a longitudinal shift of 2.55 cm (i.e. $\beta\lambda/2$ at final energy) to achieve the 180° phase shift between the beam bunches of each beam line. The measured normalized 90% RMS-emittance of the two beams are equal within 6 % [4].

## 3 FUNNELING-DEFLECTORS

To bend the beam to a common axis we use two types, the singlegap and the multigap funnel deflector.

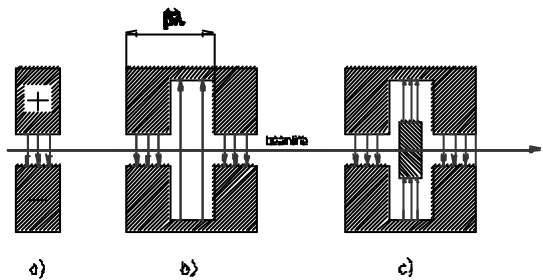

Fig. 4: Schemes of the single- and multigap funnel deflectors. The arrows show the electric field during different periods. a) a one cell singlegap deflector, b) a three cell multigap deflector, c) multigap deflector with one central drift tube.

The deflectors are mounted at the point of beam crossing, which is 52 cm behind the RFQ. This device is like a plate capacitor, oscillating with the same resonant frequency as the RFQ. The singlegap deflector is shown in Figure 5.

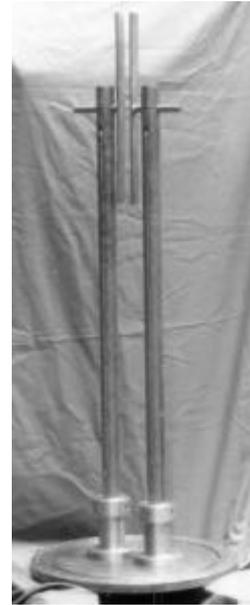

Fig. 5: Figure of the singlegap funnel deflector. The deflector discs are mounted at water-cooled stems. The length is about 2 m.

The angle between the two beam axis is 76 mrad. The singlegap funnel deflector bends this angle down to zero by a voltage, which is in our experiment about 25 kV. Figure 6 shows a simulated beam bending with the singlegap funnel deflector for one beam. The deflector bends the beam from an average angle of 38 mrad down to zero.

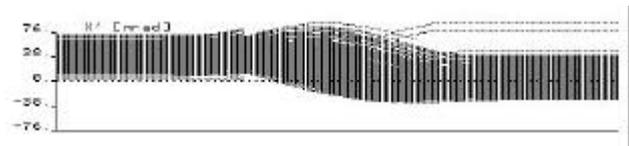

Fig. 6: Simulation of the beam bending in the singlegap funnel deflector. The angle is bending down from 38 mrad down to zero.

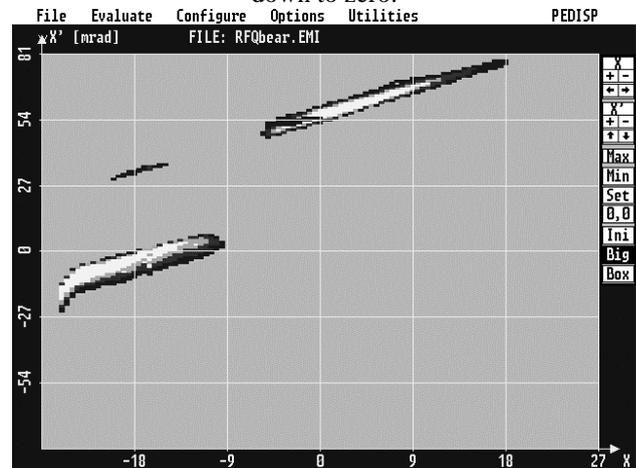

Fig. 7: Emittance measurement while the singlegap deflector is switched off.

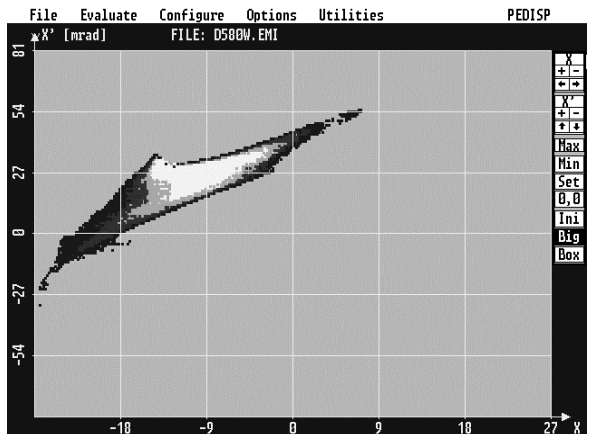

Fig. 8: Emittance measurement while the singlegap deflector is switched on. The two emittances are merged into a common beam.

Figures 7 and 8 show two emittance measurements. If the funnel deflector is switched off, the beam drifts through the deflector and we measure the beam-angle of 76 mrad. Figure 8 shows an emittance measurement with the singlegap funnel deflector switched on. The two beams are merged into a common beam.

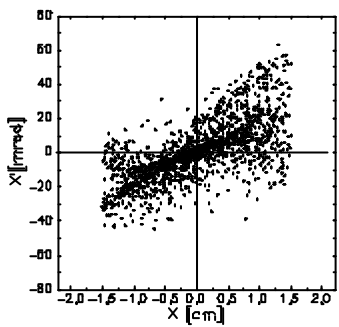

Fig. 9: Simulation of funnelling

Fig 9 shows a simulation of beam bending of two beams. The angle is reduced from 37.5 mrad down to zero. The "banana form" of the simulated emittance is caused by inhomogeneous electric fields [10].

In a multigap funnel deflector the bending is done by many gaps, which reduce the bending voltage. Figure 10 shows a beam simulation for the multigap funnel deflector with nine gaps. Figure 11 shows an emittance measurement with the multigap funnel deflector.

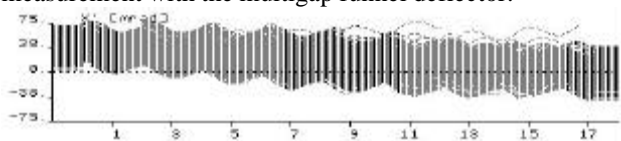

Fig. 10: Simulation of the beam bending by the multigap funnel deflector.

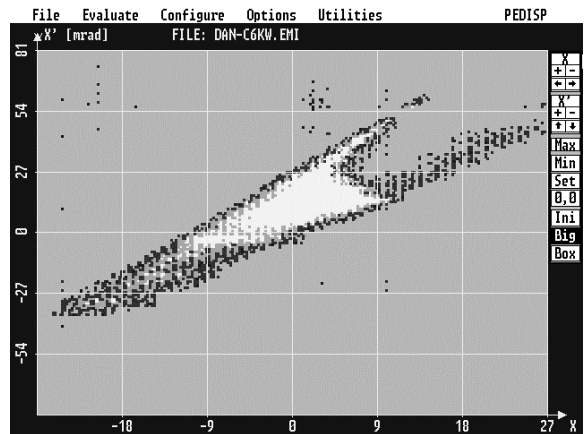

Fig. 11: Emittance measurement while the multigap deflector is switched on. The two beams are merged to a single beam.

## CONCLUSION

The measured emittances demonstrate, that both funnel deflectors brought the two beams to a common axis. The form of the measured ellipses show however, that we have to improve the matching of the beam to the RFQ to reduce the beam radius and phase width. As long as there is no matching section at the end of the RFQ electrodes, the beam is too large to measure emittance growing during funneling with our emittance measurement device.